\documentclass[twocolumn,pre]{revtex4}

\usepackage{graphicx}
\usepackage{amsmath,amssymb}

\newcommand\etal{\textit{et~al.}}
\newcommand\set[1]{\lbrace#1\rbrace}

\begin{document}

\title{Mixture models and exploratory analysis in networks}

\author{M. E. J. Newman and E. A. Leicht}
\affiliation{Department of Physics, University of Michigan, Ann Arbor,
MI 48109, USA}

\begin{abstract}
Networks are widely used in the biological, physical, and social sciences
as a concise mathematical representation of the topology of systems of
interacting components.  Understanding the structure of these networks is
one of the outstanding challenges in the study of complex systems.  Here we
describe a general technique for detecting structural features in
large-scale network data which works by dividing the nodes of a network
into classes such that the members of each class have similar patterns of
connection to other nodes.  Using the machinery of probabilistic mixture
models and the expectation-maximization algorithm, we show that it is
possible to detect, without prior knowledge of what we are looking for, a
very broad range of types of structure in networks.  We give a number of
examples demonstrating how the method can be used to shed light on the
properties of real-world networks, including social and information
networks.
\end{abstract}

\maketitle

\section{Introduction}
In the last few years, networks have found use in many fields as a powerful
tool for representing the structure of complex
systems~\cite{NBW06,AB02,Newman03d,Boccaletti06}.  Metabolic, protein
interaction, and genetic regulatory networks are now heavily studied in
biology and medicine, the Internet and the world wide web in computer and
information sciences, food webs and other species interaction networks in
ecology, and networks of personal or social contacts in epidemiology,
sociology, and the management sciences.

The study of networks goes back much further than the current surge of
interest in it, but recent work differs fundamentally from earlier studies
in the sheer scale of the networks being analyzed.  The networks studied 50
years ago by pioneers in the information and social sciences had,
typically, a few dozen vertices and were small enough that they could
easily be drawn on a piece of paper and perused for interesting features.
In the 21st century, on the other hand, networks of thousands or millions
of vertices are not unusual and network data on this scale cannot easily be
represented in a way that allows quantitative analysis to be conducted by
eye.  Instead we have been obliged to turn to topological measures,
computer algorithms, and statistics to understand the structure of today's
networks.  Much of the current research on networks is, in effect, aimed at
answering the question ``How can we tell what a network looks like, when we
can't actually look at it?''

The typical approach to this problem involves defining measures or
statistics to quantify network features of interest: centrality
indices~\cite{WF94,Scott00}, degree
distributions~\cite{AJB99,FFF99,Kleinberg99b}, clustering
coefficients~\cite{WS98}, community structure
measurements~\cite{GN02,DDDA05}, correlation
functions~\cite{PVV01,Newman02f}, and motif counts~\cite{Milo02} are all
invaluable tools for shedding light on the topology of networks.  Our
reliance on measures like these, however, has a downside: they require us
to know what we are looking for in advance before we can decide what to
measure.  People measure correlation functions, for instance, because
(presumably) they think there may be interesting correlations in a network;
they measure degree distributions because they believe the degree
distribution may show interesting features.  This approach has certainly
worked well---many illuminating discoveries have been made this way.  But
it raises an uncomfortable question: could there be interesting and
relevant structural features of networks that we have failed to find simply
because we haven't thought to measure the right thing?

To some extent this is an issue with the whole of scientific endeavor.  In
any field thinking of the right question can demand as much insight as
thinking of the answer.  However, there are also things we can do to help
ourselves.  In this paper we describe a technique that allows us to detect
structure in network data while making only rather general assumptions
about what that structure is.  Methods of this type are referred to by
statisticians as ``exploratory'' data analysis techniques, and we will make
use of a number of ideas from the statistical literature in the
developments that follow.

We focus on the problem of classifying the vertices of a network into
groups such that the members of each group are similar in some sense.  This
already narrows the types of structure we consider substantially, but
leaves a large and useful selection of types still in play.  Some of these
types have been studied in the past, but the range of possibilities
considered here is far larger than that of previous work.  For instance,
many researchers have examined ``community structure'' in networks---also
called ``homophily'' or ``assortative mixing''---in which vertices divide
into groups such that the members of each group are mostly connected to
other members of the same group~\cite{GN02,DDDA05}.  ``Disassortative
mixing,'' in which vertices have most of their connections outside their
group, has also been discussed to a lesser
extent~\cite{HLEK03,Newman06c,ER05}.  Effective techniques have been
developed that can detect structure of both of these types.  But what
should we do if we do not know in advance which type to expect or if our
network has some other type of structure entirely whose existence we are
not even aware of?  One can imagine an arbitrary number of other types of
division among the vertices of a network, most of which have probably never
been considered explicitly in the past.  One possibility, for instance, is
a network in which, although there is no conventional assortative mixing,
there are certain ``keystone'' vertices and group membership is defined by
which particular keystone or set of keystones a vertex is connected to.
Another possibility is a network in which there is both assortative and
disassortative mixing between members of the same groups, the groups
themselves being defined by the fact that their vertices have the same
pattern of preferences and aversions, rather than by any overall
assortative or disassortative behavior at the group level.  And there are
certainly many other possibilities.  Such complex structures cannot be
detected by the standard methods available to us at present and moreover it
seems unlikely in many cases that appropriate specialized detection methods
will be developed because of the chicken-and-egg nature of the problem: we
would have to know the form of the structure in question to develop such a
method, but without a detection method we cannot discover that form in the
first place.

Here we propose a new approach to the structural analysis of network data
that aims to circumvent these issues.  It does so by employing a broad and
flexible definition of vertex classes, parametrized by an extensive number
of variables and hence encompassing an essentially infinite variety of
structural types in the limit of large network size.  Certainly our
definition includes the standard assortative and disassortative structures
discussed above and, as we will see, the method we propose will detect
those structures when they are present.  But it is also able to detect a
wide variety of other structural types and, crucially, does so without
requiring us to specify in advance which particular structure we are
looking for: the method simultaneously finds the appropriate assignment of
vertices to groups and the parameters defining the meaning of those groups,
so that upon completion the calculation tells us not only the best way of
grouping the vertices but also the definitions of the groups themselves.
Our method, which is based on the numerical technique known as the
expectation-maximization algorithm, is also fast and simple to implement.
We demonstrate the algorithm with applications to a selection of real-world
networks and computer-generated test networks.

\section{The method}
The method we describe is based on a mixture model, a standard construct in
statistics, though one that has not yet found wide use in studies of
networks.  The method works well for both directed and undirected networks,
but is somewhat simpler in the directed case, so let us start there.

Suppose we have a network of $n$ vertices connected by directed edges, such
as a web graph or a food web.  The network can be represented
mathematically by an adjacency matrix~$A$ with elements $A_{ij} = 1$ if
there is an edge from $i$ to~$j$ and 0 otherwise.

Suppose also that the vertices fall into some number~$c$ of classes or
groups and let us denote by $g_i$ the group to which vertex~$i$ belongs.
We will assume that these group memberships are unknown to us and that we
cannot measure them directly.  In the language of statistical inference
they are ``hidden'' or ``missing'' data.  Our goal is to infer them from
the observed network structure.  (The number of groups~$c$ can also be
inferred from the data, but for the moment we will treat it as given.)  To
infer the group memberships we adopt a standard approach for such problems:
we propose a flexible (mixture) model for the groups and their properties,
then vary the parameters of the model into order to find the best fit to
the observed network.

The model we use is a stochastic one that parametrizes the probability of
each possible configuration of group assignments and edges as follows.  We
define $\theta_{ri}$ to be the probability that a (directed) link from a
particular vertex in group~$r$ connects to vertex~$i$.  In the world wide
web, for instance, $\theta_{ri}$~would represent the probability that a
hyperlink from a web page in group~$r$ links to web page~$i$.  In effect
$\theta_{ri}$ represents the ``preferences'' of vertices in group~$r$ about
which other vertices they link to.  In our approach it is these preferences
that define the groups: a ``group'' is a set of vertices that all have
similar patterns of connection to others~\cite{note1}.  (The idea is
similar in philosophy to the block models proposed by White and others for
the analysis of social networks~\cite{WBB76}, although the realization and
the mathematical techniques employed are different.)  Note that there is no
assumption that the vertices~$i$ to which the members of a group link
themselves belong to any particular group or groups; they can be in the
same group or in different groups or randomly distributed over the entire
network.  Thus the structures we envisage can be quite different from
traditional assortatively mixed networks, although they include the latter
as a special case.

We also define $\pi_r$ be the (currently unknown) fraction of vertices in
group or class~$r$, or equivalently the probability that a randomly chosen
vertex falls in~$r$.  The parameters~$\pi_r,\theta_{ri}$ satisfy the
normalization conditions
\begin{equation}
\sum_{r=1}^c \pi_r = 1,\qquad \sum_{i=1}^n \theta_{ri} = 1.
\label{eq:normalization}
\end{equation}

The quantities in our theory thus fall into three classes: measured
data~$\set{A_{ij}}$, missing data~$\set{g_i}$, and model
parameters~$\set{\pi_r,\theta_{ri}}$.  To simplify the notation we will
henceforth denote by $A$ the entire set $\set{A_{ij}}$ and similarly for
$\set{g_i}$, $\set{\pi_r}$, and~$\set{\theta_{ri}}$.

The standard framework for fitting models like this one to a given data set
is likelihood maximization, in which one maximizes with respect to the
model parameters the probability that the data were generated by the given
model.  Maximum likelihood methods have occasionally been employed in
network calculations in the past~\cite{JH03b,CNM06,Hastings06}, as well as
in many other problems in the study of complex systems more generally.  In
the present case, our fitting problem requires us to maximize the
likelihood $\Pr(A,g|\pi,\theta)$ with respect to $\pi$ and~$\theta$, which
can be done by writing
\begin{equation}
\Pr(A,g|\pi,\theta) = \Pr(A|g,\pi,\theta)\Pr(g|\pi,\theta),
\end{equation}
where
\begin{equation}
\Pr(A|g,\pi,\theta) = \prod_{ij} \theta_{g_i,j}^{A_{ij}},\quad
\Pr(g|\pi,\theta) = \prod_i \pi_{g_i},
\label{eq:pra}
\end{equation}
so that the likelihood is
\begin{equation}
\Pr(A,g|\pi,\theta) = \prod_i \biggl[ \pi_{g_i} \prod_j
                        \theta_{g_i,j}^{A_{ij}} \biggr].
\label{eq:llag}
\end{equation}
In fact, one commonly works not with the likelihood itself but with its
logarithm:
\begin{equation}
\mathcal{L} = \ln \Pr(A,g|\pi,\theta)
            = \sum_i \Bigl[ \ln \pi_{g_i} +
              \sum_j A_{ij} \ln\theta_{g_i,j} \Bigr].
\label{eq:condll}
\end{equation}
The maximum of the two functions is in the same place, since the logarithm
is a monotonically increasing function.

Unfortunately, $g$~is unknown in our case, which means the value of the
log-likelihood is also unknown.  We can, however, usually make a good guess
at the value of~$g$ given the network structure~$A$ and the model
parameters~$\pi,\theta$.  More specifically we can, as shown below,
calculate the probability distribution $\Pr(g|A,\pi,\theta)$ and from it
calculate an expected value~$\overline{\mathcal{L}}$ for the log-likelihood
by averaging over~$g$ thus:
\begin{eqnarray}
\overline{\mathcal{L}} &=& \sum_{g_1=1}^c\!\ldots\!\!\sum_{g_n=1}^c
      \Pr(g|A,\pi,\theta) \sum_i \Bigl[ \ln \pi_{g_i} +
      \sum_j A_{ij} \ln\theta_{g_i,j} \Bigr] \nonumber\\
  &=& \sum_{ir} \Pr(g_i=r|A,\pi,\theta) \Bigl[ \ln \pi_r +
      \sum_j A_{ij} \ln\theta_{rj} \Bigr] \nonumber\\
  &=& \sum_{ir} q_{ir} \Bigl[ \ln \pi_r +
      \sum_j A_{ij} \ln\theta_{rj} \Bigr],
\label{eq:ll}
\end{eqnarray}
where to simplify the notation we have defined
$q_{ir}=\Pr(g_i=r|A,\pi,\theta)$, which is the probability that vertex~$i$
is a member of group~$r$.  (In fact, it is precisely these probabilities
that will be the principal output of our calculation.)

This expected log-likelihood represents our best estimate of the value
of~$\mathcal{L}$ and the position of its maximum represents our best
estimate of the most likely values of the model parameters.  Finding the
maximum still presents a problem, however, since the calculation of~$q$
requires the values of $\pi$ and $\theta$ and the calculation of $\pi$ and
$\theta$ requires~$q$.  The solution is to adopt an iterative,
self-consistent approach that evaluates both simultaneously.  This type of
approach, known as an expectation-maximization or EM algorithm, is common
in the literature on missing data problems.  In its modern form it is
usually attributed to Dempster~\etal~\cite{DLR77}, who built on theoretical
foundations laid previously by a number of other authors~\cite{MK96}.

Following the conventional development of the method, we calculate the
expected probabilities~$q$ of the group memberships given $\pi,\theta$ and
the observed data thus:
\begin{equation}
q_{ir} = \Pr(g_i=r|A,\pi,\theta)
       = {\Pr(A,g_i=r|\pi,\theta)\over\Pr(A|\pi,\theta)}.
\label{eq:qbar1}
\end{equation}
The factors on the right are given by summing over the possible values
of~$g$ in Eq.~\eqref{eq:llag} thus:
\begin{eqnarray}
\Pr(A,g_i=r|\pi,\theta) &=& \sum_{g_1=1}^c\ldots\sum_{g_n=1}^c
      \delta_{g_i,r} \Pr(A,g|\pi,\theta) \nonumber\\
  &=& \sum_{g_1=1}^c\ldots\sum_{g_n=1}^c \delta_{g_i,r}
      \prod_k \biggl[ \pi_{g_k} \prod_j \theta_{g_k,j}^{A_{kj}} \biggr]
      \nonumber\\
  &=& \biggl[ \pi_r \prod_j \theta_{rj}^{A_{ij}} \biggr]
      \biggl[ \prod_{k\ne i} \sum_{s=1}^c \pi_s \prod_j \theta_{sj}^{A_{kj}}
      \biggr], \nonumber\\
\end{eqnarray}
and
\begin{eqnarray}
\Pr(A|\pi,\theta) &=& \sum_{g_1=1}^c\ldots\sum_{g_n=1}^c \Pr(A,g|\pi,\theta)
                      \nonumber\\
                  &=& \prod_k \sum_{s=1}^c \pi_s \prod_j \theta_{sj}^{A_{kj}},
\end{eqnarray}
where $\delta_{ij}$ is the Kronecker $\delta$ symbol.  Substituting
into~\eqref{eq:qbar1}, we then find
\begin{equation}
q_{ir} = {\pi_r \prod_j \theta_{rj}^{A_{ij}}\over
                \sum_s \pi_s \prod_j \theta_{sj}^{A_{ij}}}.
\label{eq:estep}
\end{equation}
Note that $q_{ir}$ correctly satisfies the normalization condition
$\sum_r q_{ir} = 1$.

Once we have the values of the~$q_{ir}$, we can use them to evaluate the
expected log-likelihood, Eq.~\eqref{eq:ll}, and hence to find the values of
$\pi,\theta$ that maximize it.  One advantage of the current approach now
becomes clear: because the~$q_{ir}$ are known, fixed quantities, the
maximization can be carried out purely analytically, obviating the need for
numerical techniques such as Markov chain Monte Carlo.  Introducing
Lagrange multipliers to enforce the normalization conditions,
Eq.~\eqref{eq:normalization}, and differentiating, we find that the maximum
of the likelihood occurs when
\begin{equation}
\pi_r = {1\over n} \sum_i q_{ir},\qquad
\theta_{rj} = {\sum_i A_{ij} q_{ir}\over
               \sum_i k_i q_{ir}},
\label{eq:mstep}
\end{equation}
where $k_i = \sum_j A_{ij}$ is the out-degree of vertex~$i$ and we have
explicitly evaluated the Lagrange multipliers using the normalization
conditions.

Equations~\eqref{eq:estep} and \eqref{eq:mstep} define our
expectation-maximization algorithm.  Implementation of the algorithm
consists merely of iterating these equations to convergence and the output
is the probability~$q_{ir}$ for each vertex to belong to each group, plus
the probabilities $\theta_{ri}$ of links from vertices in each group to
every other vertex, the latter effectively giving the definitions of the
groups.  The calculation converges rapidly in practice: typical run times
for the networks studied here were fractions of a second.  (Some
theoretical results are known for the convergence of EM algorithms---see
Dempster~\etal~\cite{DLR77} and Wu~\cite{WU83}.)

The obvious choice of starting values for the iteration is the symmetric
choice $\pi_i=1/c$, $\theta_{ri}=1/n$, but unfortunately these values are a
trivial (unstable) fixed point of Eqs.~\eqref{eq:estep} and
\eqref{eq:mstep} and hence a poor choice.  In our calculations we have
instead used starting values that are perturbed randomly a small distance
from this fixed point.  A random starting condition also gives us an
opportunity to assess the robustness of our results.  Except in special
cases (such as the trivial fixed point above), expectation-maximization
algorithms are known to converge to local maxima of the
likelihood~\cite{MK96} but not always to global maxima, and hence it is
possible to get different solutions from different starting points.  The
method works well in cases where it frequently converges to the global
maximum or where it converges to local maxima that are close to the global
maximum, giving good if not perfect solutions on most runs.  In practice,
we find for some networks that the method almost always converges to the
same solution or a very similar one, while for others it is necessary to
perform several runs with different initial conditions to find a good
maximum of the likelihood.  In the calculations presented in this paper, we
have in each case taken the division of the network giving the highest
likelihood over the runs performed.

The developments so far apply to the case of a directed network.  Most of
the networks studied in the recent literature, however, are undirected.
The model used above is inappropriate for the undirected case because its
edges represent an inherently asymmetric, directed relationship between
vertices in which one vertex chooses unilaterally to link to another, the
receiving vertex having no say in the matter.  The edges in an undirected
network, by contrast, usually represent symmetric relationships.  In a
social networks of friendships, for instance, the edges would typically be
drawn undirected because two people can become friends only if both choose
to be friendly towards the other.  To extend our method to undirected
networks we need to incorporate this symmetry into our model, which we do
as follows.  Once again we define $\theta_{ri}$ to be the probability that
a vertex in group~$r$ ``chooses'' to link to vertex~$i$, but we now specify
that a link will be formed only if two vertices both choose each other.
Thus the probability that an edge falls between vertices~$i$ and~$j$, given
that $i$ is in group~$s$ and $j$ is in group~$r$,
is~$\theta_{ri}\theta_{sj}$, which is now symmetric.  This probability
satisfies the normalization condition $\sum_{ij} \theta_{ri}\theta_{sj}=1$
for all $r,s$ and setting $r=s$ we find
\begin{equation}
\sum_{ij} \theta_{ri}\theta_{rj} = \Bigl[ \sum_i \theta_{ri} \Bigr]^2 =
1,
\end{equation}
and hence $\sum_i \theta_{ri}=1$ as before.

Now the probability $\Pr(A|g,\pi,\theta)$ in Eq.~\eqref{eq:llag} is given
by
\begin{equation}
\Pr(A|g,\pi,\theta) = \prod_{i>j} \bigl[ \theta_{g_i,j}\theta_{g_j,i}
                      \bigr]^{A_{ij}}
                    = \prod_{ij} \theta_{g_i,j}^{A_{ij}},
\end{equation}
exactly as in the directed case, where we have made use of the fact that
$A_{ji}=A_{ij}$ for an undirected network.  (We have also assumed there are
no self-edges in the network---edges that connect a vertex to itself---so
that $A_{ii}=0$ for all~$i$.)

The remainder of the derivation now follows as before and results in the
same equations, \eqref{eq:estep}~and~\eqref{eq:mstep}, for the final
algorithm.

\section{Example applications}
For our first examples of the operation of our method, we apply it to two
small networks, one known to have conventional assortative community
structure, the other known to have disassortative structure.  The first is
the much-discussed ``karate club'' network of friendships between 34
members of a karate club at a US university, assembled by
Zachary~\cite{Zachary77} by direct observation of the club's members.  This
network is of particular interest because the club split in two during the
course of Zachary's observations as a result of an internal dispute and
Zachary recorded the membership of the two factions after the split.

\begin{figure}
\includegraphics[width=8cm]{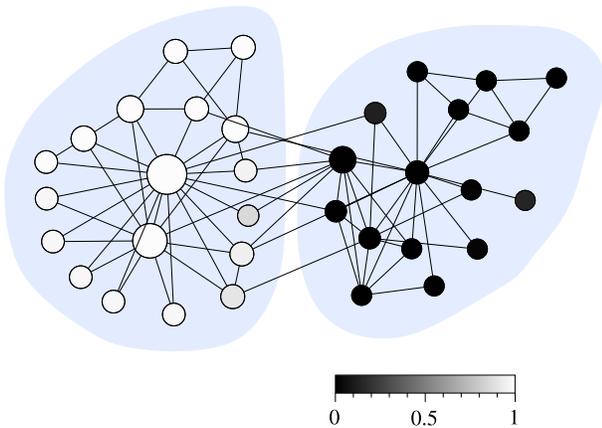}
\caption{Application of the method described here to the ``karate club''
  network of Ref.~\cite{Zachary77}.  The two shaded regions indicate the
  division of the network in the real world, while the shades of the
  individual vertices indicate the decomposition chosen by the algorithm.
  The sizes of the vertices indicate the probabilities $\theta_{1i}$ for
  edges from vertices in group~1 (the left-hand group) to be connected to
  each other vertex, with the probabilities ranging from 0 for the smallest
  vertices to 0.19 for the largest.}
\label{fig:zachary}
\end{figure}

Figure~\ref{fig:zachary} shows the best division of this network into two
groups found using the expectation-maximization method with $c$ set equal
to~2.  The shades of the vertices in the figure represent the values of the
variables $q_{i1}$ for each vertex on the scale shown (or equivalently the
values of $q_{i2}$, since $q_{i1}+q_{i2}=1$ for all~$i$).  As we can see
the algorithm assigns most of the vertices strongly to one group or the
other; in fact, all but 13 vertices are assigned 100\% to one of the groups
(black and white vertices in the figure).  Thus the algorithm finds a
strong split into two clusters in this case, and indeed if one simply
divides the vertices according to the cluster to which each is most
strongly assigned, the result corresponds perfectly to the division
observed in real life (denoted by the shaded regions in the figure).

But the algorithm reveals much more about the network than this.  First,
where appropriate it can return probabilities for assignment to the two
groups that are not 0 or 1 but lie somewhere between these limits, and for
some of the vertices in this network it does so.  Inspection of the figure
reveals in particular a small number of vertices with intermediate shades
of gray along the border between the groups.  There has been some
discussion in the recent literature of methods for divining ``fuzzy'' or
overlapping groups in networks; rather than dividing a network sharply into
groups, it is sometimes desirable to assign vertices to more than one group
and a number of possible ways of doing this have been
proposed~\cite{RB04,PDFV05,BGM05,Newman06c}.  The present algorithm offers
an alternative method that is particularly attractive because of the clear
definition of the overlap: the values of the $q_{ir}$ give the precise
probability that a vertex belongs to a specified group, given the observed
network structure.

The algorithm also returns the distributions or preferences~$\theta_{ri}$
for connections from vertices in group~$r$ to each other vertex~$i$.  In
Fig.~\ref{fig:zachary} we indicate by the sizes of vertices the
probabilities~$\theta_{1i}$ of edges from vertices in group~1, which is the
left-hand group in the figure, to connect to each other vertex.  As we can
see, two vertices central to the group have high connection probabilities,
while some of the more peripheral vertices have smaller probabilities.
Thus the values of $\theta_{ri}$ behave as a kind of centrality measure,
indicating how important a particular vertex is to a particular group.
This could form the basis for a practical measure of within-group influence
or attraction in social or other networks.  Note that in this case this
measure is not high for vertices that are central to the other group,
group~2; the measure is sensitive to the particular preferences of the
vertices in just a single group.

\begin{figure}
\includegraphics[width=8cm]{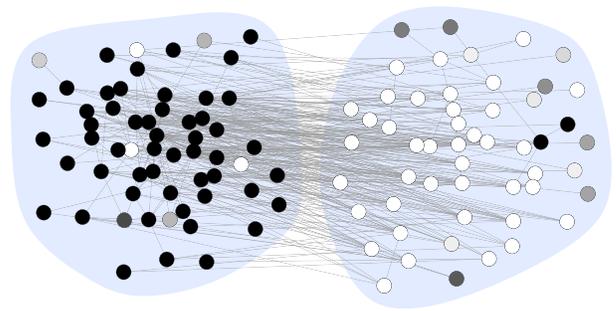}
\caption{The adjacency network of English words described in the text.  The
two shaded groups contain adjectives and nouns respectively and the shades
of the individual vertices represent the classification found by the
algorithm.}
\label{fig:words}
\end{figure}

We can take the method further.  In Fig.~\ref{fig:words} we show the
results of its application to an adjacency network of English words taken
from Ref.~\cite{Newman06c}.  In this network the vertices represent 112
commonly occurring adjectives and nouns in a particular body of text (the
novel \textit{David Copperfield} by Charles Dickens), with edges connecting
any pair of words that appear adjacent to each other at any point in the
text.  Since adjectives typically occur next to nouns in English, most
edges connect an adjective to a noun and the network is thus approximately
bipartite or disassortative.  This can be seen clearly in the figure, where
the two shaded groups represent the adjectives and nouns and most edges are
observed to run between groups.

Analyzing this network using our algorithm we find the classification shown
by the shades of the vertices.  Once again most vertices are assigned 100\%
to one class or the other, although there are a few ambiguous cases,
visible as the intermediate shades.  As the figure makes clear, the
algorithm's classification corresponds closely to the adjective/noun
division of the words---almost all the black vertices are in one group and
the white ones in the other.  In fact, 89\% of the vertices are correctly
classified by our algorithm in this case.

The crucial point to notice, however, is that the algorithm is not merely
able to detect the bipartite structure in this network, but it is able to
do so without being told that it is to look for bipartite structure.  The
exact same algorithm, unmodified, finds both the assortative structure of
Fig.~\ref{fig:zachary} and the disassortative structure of
Fig.~\ref{fig:words}.  This is an important strength of the present method:
it is able to detect a range of different structural types without knowing
in advance what type to expect.  Other methods are able to detect
particular kinds of structure, and in many cases do a good job, but they
tend to be narrowly tailored to that job---typically a new method or
algorithm has to be devised for each new structural type.

\begin{figure}
\includegraphics[width=8cm]{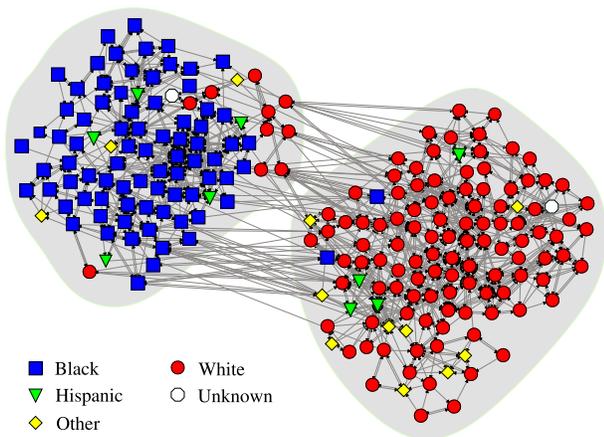}
\caption{A directed social network of US high school students and the
  division into two groups found by the directed version of our method.
  Vertex shapes show the (self-identified) ethnicity of the students, as
  indicated.}
\label{fig:addhealth}
\end{figure}

The networks in Figs.~\ref{fig:zachary} and~\ref{fig:words} are both
undirected, but our method is applicable to directed networks as well.  In
Fig.~\ref{fig:addhealth} we show an example of a directed network, a social
network of high school students taken from the US National Longitudinal
Study of Adolescent Health (the ``AddHealth'' study)~\cite{note2}.
Students were asked to identify their friends within the school and a
response in which student~A identifies B as a friend is represented as a
directed edge from A to~B.  In contrast to the common view, discussed
earlier, of friendship as a symmetric relationship running in both
directions between the individuals it connects, a remarkable number of the
friendships identified in this study---more than half---are found to run in
only one direction, so that a directed representation of the network is
indispensable for capturing the structure of the data.

Applying the directed version of our method to this network with $c=2$
produces the division shown in Fig.~\ref{fig:addhealth}.  This example is
striking because, like many of the networks in the AddHealth data set, the
groupings are found to correlate strongly with student ethnicity as shown
by the shapes of the vertices~\cite{Moody01}.  In this case, one of the two
groups contains most of the school's black students and the other most of
the white students, with the few members of other ethnic groups distributed
more evenly.

\begin{figure*}
\includegraphics[width=14cm]{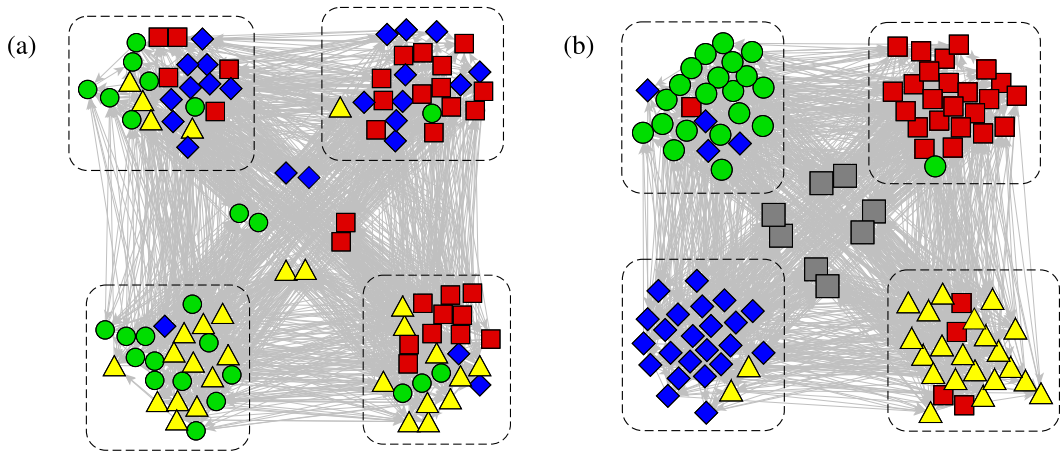}
\caption{The four-group network described in the text, in which connections
  between vertices are entirely random, except for connections to the eight
  keystone vertices in the center.  Each of the four groups (dashed boxes)
  is thus distinguished solely by the unique pattern of its connections to
  the keystone vertices.  Vertex shapes represent the groups to which
  vertices are assigned by our analyses using (a)~the traditional community
  detection and (b)~maximum likelihood methods of this paper.}
\label{fig:keystone}
\end{figure*}

The examples we have seen so far all center on networks with strong
assortative or disassortative mixing, but it is important to emphasize that
our method is applicable to other types of structure as well.  For our
final example we focus on a network of a completely different kind, a
computer-generated network of a form mentioned in the introduction.  In
this network there are a small number of ``keystone'' vertices and group
membership affects only the propensity to link to these vertices; all other
connections are purely random.  In detail the network is as follows.

The network is again a directed one, with a total of 108 vertices.  100 of
those vertices are divided into four groups of 25 each and directed edges
are placed uniformly at random between them such that the mean degree (both
in and out) is ten.  The remaining eight vertices are denoted keystone
vertices and the other vertices link to them depending on their group
membership.  Specifically, the vertices in groups~A, B, C, and D link to
keystone vertices $\set{1,2,3,4}$, $\set{3,4,5,6}$, $\set{5,6,7,8}$, and
$\set{7,8,1,2}$ respectively.  Thus no keystone vertex is uniquely
identified with any group, but each group has a unique signature set of
keystones and it is only this pattern of keystone links that distinguishes
the group.  The network is not assortative (or disassortative) by the
traditional definition: the randomly placed edges fall within or between
groups purely according to chance, and the links to the keystones, while
not random, are equally likely to fall within or between groups.

Figure~\ref{fig:keystone}a shows what happens when we analyze this network
using a standard community detection technique.  The dashed boxes in the
figure outline the four groups of vertices and the shapes show the group
assignments found by the analysis.  Although the analysis does find four
groups in this case, the groups do not correspond to the known division of
the network---each box contains substantial numbers of vertices of at least
two types and in some cases more.  The maximum likelihood analysis, by
contrast, has no difficulty in discerning the structure of the network.
Figure~\ref{fig:keystone}b shows the output of our algorithm with $c=4$
and, as we can see, the algorithm has, without prior knowledge of the type
of structure in the network, discovered the structure and correctly
assigned almost all of the vertices to their groups.  The eight keystone
vertices, which are shown in the center of the panel, are not assigned to
any group by the algorithm, but are instead divided (almost) equally
between all four (meaning that $q_{ir}$ is close to 0.25 for all~$r$).
Thus the algorithm has, in effect, accurately deduced the five classes of
vertices present in the network.  Moreover an examination of the final
values of the model parameters $\theta_{ri}$ will tell us exactly what type
of structure the algorithm has discovered.  In principle, considerably more
complex structures than this can be detected as well.

\section{Conclusions}
In this paper we have described a method for exploratory analysis of
network data in which vertices are classified into groups based on the
observed patterns of connections between them.  The method is more general
than previous clustering methods, making use of maximum likelihood
techniques to classify vertices and simultaneously determine the definitive
properties of each class.  The result is a simple algorithm that is capable
of detecting a broad range of structural signatures in networks, including
conventional community structure, bipartite or disassortative structure,
fuzzy or overlapping classifications, and many mixed or hybrid structural
forms that have not been considered explicitly in the past.  We have
demonstrated the method with applications to a variety of examples,
including real-world and computer-generated networks.  The method's
strength is its flexibility, which will allow researchers to probe observed
networks for general types of structure without having to specify in
advance what type they expect to find.

\begin{acknowledgments}
The authors thank Marian Bogu\~na, Aaron Clauset, Carrie Ferrario,
Cristopher Moore, and Kerby Shedden for useful conversations and comments
and Northwestern University for hospitality and support while part of this
work was conducted.  The work was funded in part by the National Science
Foundation under grant number DMS--0405348 and by the James S. McDonnell
Foundation.
\end{acknowledgments}


\begin{thebibliography}{10}
\bibitem{note1}
  We could alternatively base our calculation on the patterns of ingoing
  rather than outgoing links and for some networks this may be a useful
  approach.  The mathematical developments are entirely analogous to those
  presented here.
\end{thebibliography}
\end{document}